\documentclass[aip,reprint]{revtex4-2}

\usepackage{epsfig,graphicx,times}
\usepackage{bm}
\usepackage[utf8]{inputenc}
\usepackage[T1]{fontenc}
\usepackage{mathptmx}
\usepackage{etoolbox}
\usepackage{ulem}
\usepackage{color}
\usepackage{amstext}
\usepackage{amsmath}            
\usepackage{amssymb}            
\usepackage{graphicx}           
\usepackage{latexsym}

\begin{document}
\title{Fabrication of airbridges with gradient exposure}
\author{Yuting Sun}
\email[]{qu_syt@163.com}
\affiliation{National Laboratory of Solid State Microstructures, School of Physics, Nanjing University, Nanjing 210093, China}
\affiliation{Hefei National Laboratory, Hefei 230088, China}
\author{Jiayu Ding}
\affiliation{National Laboratory of Solid State Microstructures, School of Physics, Nanjing University, Nanjing 210093, China}
\affiliation{Hefei National Laboratory, Hefei 230088, China}
\author{Xiaoyu Xia}
\affiliation{National Laboratory of Solid State Microstructures, School of Physics, Nanjing University, Nanjing 210093, China}
\affiliation{Hefei National Laboratory, Hefei 230088, China}
\author{Xiaohan Wang}%
\affiliation{National Laboratory of Solid State Microstructures, School of Physics, Nanjing University, Nanjing 210093, China}
\affiliation{Hefei National Laboratory, Hefei 230088, China}
\author{Jianwen Xu}
\affiliation{National Laboratory of Solid State Microstructures, School of Physics, Nanjing University, Nanjing 210093, China}
\affiliation{Hefei National Laboratory, Hefei 230088, China}
\author{Shuqing Song}
\affiliation{National Laboratory of Solid State Microstructures, School of Physics, Nanjing University, Nanjing 210093, China}
\affiliation{Hefei National Laboratory, Hefei 230088, China}
\author{Dong Lan}
\affiliation{National Laboratory of Solid State Microstructures, School of Physics, Nanjing University, Nanjing 210093, China}
\affiliation{Hefei National Laboratory, Hefei 230088, China}
\author{Jie Zhao}
\email[]{jiezhao@nju.edu.cn}
\affiliation{National Laboratory of Solid State Microstructures, School of Physics, Nanjing University, Nanjing 210093, China}
\affiliation{Hefei National Laboratory, Hefei 230088, China}
\affiliation{Shishan Laboratory, Suzhou Campus of Nanjing University, Suzhou 215000, China}
\author{Yang Yu}
\affiliation{National Laboratory of Solid State Microstructures, School of Physics, Nanjing University, Nanjing 210093, China}
\affiliation{Hefei National Laboratory, Hefei 230088, China}

\date{August 1 2022}

\begin{abstract}
In superconducting quantum circuits, airbridges are critical for eliminating parasitic slotline modes of coplanar waveguide circuits and reducing crosstalks between direct current magnetic flux biases.
Here, we present a technique for fabricating superconducting airbridges. With this technique, a single layer of photoresist is employed, and the gradient exposure process is used to define the profile of airbridges. In order to properly obtain the bridge profile, we design exposure dosage based on photoresist thickness reduction and laser power calibrations. Compared with other airbridge fabrication techniques, the gradient exposure fabrication technique provides the ability to produce lossless superconducting airbridges with flexible size and thus is more suitable for large-scale superconducting quantum circuits. Furthermore, this method reduces the complexity of the fabrication process and provides a high fabrication yield.

\end{abstract}

\maketitle

Superconducting quantum circuits are one of the most promising candidates for quantum computation. Superconducting coplanar waveguide (CPW) transmission lines and resonators are critical components of such quantum circuits. To achieve specific circuit layouts, the continuities and symmetry of the finite-ground CPWs are usually sacrificed. Ground discontinuities and asymmetries may lead to parasitic modes of the circuits, which can couple to the superconducting qubits or other resonant modes of the circuits.~\cite{1991_discontinuities,qubit_modes} This coupling may lead to complex dynamical behaviors, or even reduce of the qubit coherence time.
In addition, increased manipulation channels in quantum circuits exacerbate the ground discontinuities of CPWs, which may introduce crosstalks between direct current (DC) biases of qubits. To eliminate parasitic modes, metallic crossovers are fabricated to carry out electric connections between the discrete ground planes.~\cite{1996slotline_book,slotline2,slotline3}
A well-established electric connection between discontinuous grounds of CPWs can also suppress crosstalks and facilitate quantum operations on multi-qubit superconducting quantum chips.~\cite{McDermott2005, chen2014fabrication,Jin}

To achieve this goal, several types of electric connections between discrete CPW grounds, such as wirebonds, dielectric bridges, and airbridges have been proposed and fabricated. However, the large size and non-negligible parasitic inductance of wirebonds limit their applications in superconducting quantum computation circuits.~\cite{rosa_1908,wirebond_crosstalk,chen2014fabrication} The dielectric loss and additional shunting capacitance introduced by dielectric bridges hinder the improvement of quantum coherence time of superconducting circuits.~\cite{parasitic_capacitance,Jin2021,Jin}
In order to circumvent these weaknesses, airbridges are introduced to implement the electric connections between discontinuous CPW grounds. Traditionally, airbridges are fabricated based on the e-beam lithography (EBL) technique~\cite{simple_EBL,multiple_EBL} or the re-flow followed photolithography technique~\cite{reflow_AB,chen2014fabrication}. Due to the limited thickness of EBL resist, the height and length of airbridges fabricated with the EBL technique are usually much smaller than those produced by the re-flow followed photolithography technique. The re-flow followed photolithography technique produces arch-bridge-shaped airbridges, which are higher and longer. With this fabrication technique, a bake temperature around 150$^{\circ}$C is necessary to form the arch bridge shape, and the metal connection is fabricated by subsequent deposition and etching.~\cite{photoresist} However, the high temperature used in the re-flow process leads to the formation of amorphous interlayers ($\rm \alpha$ interlayer) in the silicon-niobium interface or the silicon-tantalum interface.~\cite{SiNb_inter} This amorphous interlayer can induce dielectric loss and seriously reduce the coherent performance of quantum circuits.~\cite{interlayer_TLS} The high-temperature process may also change the critical current of Josephson junctions and thus alter the qubit spectra.~\cite{anneal}

\begin{figure}[hptb]
\includegraphics[width=\linewidth]{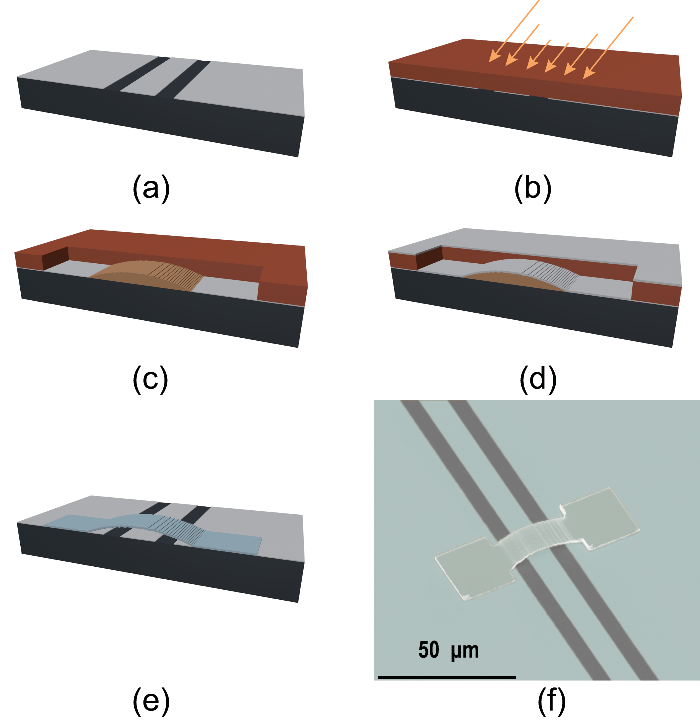}
\caption{\label{fig:fig1} (a)-(e) The gradient exposure fabrication process: (a) The wafer with pre-fabricated CPW transmission line. (b) The wafer is coated with photoresist and exposed. The different lengths of the orange arrows represent different applied exposure dosage. (c) The gradient exposure result after development. The dark red part represents the unexposed photoresist, and the dark yellow part represents the gradient exposed area. (d) The schematic after Al deposition. (e) The airbridge after liftoff. (f) SEM image of the fabricated airbridge.}
\end{figure}

Here, we introduce an airbridge fabrication technology that is suitable for superconducting quantum circuits. Our scheme employs the gradient exposure technique to form the arch bridge profile. In order to obtain the proper bridge shape, we first calibrate the relationship between exposure dosage and residual photoresist thickness with different substrate materials. Based on this calibration, we can create an airbridge design with proper exposure dosage. Using a laser lithography system and proper development process, we transform airbridges with an arch bridge profile to photoresist with thickness in micrometers. Followed by e-beam evaporation and a liftoff process, sizeable airbridges with micro-meter heights can be obtained. Compared with the EBL airbridge fabrication technique, our process can produce airbridges that are more suitable for superconducting quantum circuits. At the same time, the high temperature process used in the re-flow followed photolithography technique is avoided, so corresponding negative effects are circumvented.

The fabrication process is shown schematically in  Fig.~\ref{fig:fig1}(a) - (e). A CPW transmission line with a $10\ \rm \mu m$ wide center conductor and $5\ \rm \mu m$ wide gaps is pre-fabricated on a silicon or sapphire substrate, as shown in Fig.~\ref{fig:fig1} (a). First, a layer of Megaposit SPR-220-3 photoresist is spun onto the pre-patterned wafer with a spin speed of $1500\ \rm rpm$. Then the wafer is baked at $95^{\ \circ} \rm C$ for $150 \rm s$, and the thickness of the photoresist is about $4\ \rm \mu m$, as shown in Fig.~\ref{fig:fig1} (b). Next, based on the exposure parameters obtained in the dosage calibration process, the designed pattern is exposed using a Heidelberg DWL66+ laser lithography system. After developing the sample with the tetramethyl ammonium hydroxide based developer, we obtain the arch bridge profile in the exposed area. To facilitate the subsequent liftoff process, we design the top of the arch bridge to be about $1\ \rm \mu m$ lower than the unexposed area, as shown in Fig.~\ref{fig:fig1} (c). The varying thickness of the developed photoresist is achieved by changing the exposure dosage step by step.

After development, the wafer is transferred into an ultrahigh vacuum (UHV) e-beam evaporator for aluminum deposition. In the evaporator, an \begin{itshape}in situ\end{itshape} argon-ion milling process is applied to remove the aluminum oxidation layer on top of aluminum films, and thus realize reliable electric contact between CPW grounds and airbridges. Then, an aluminum layer of $500 \rm nm$ thickness is deposited, as shown in Fig.~\ref{fig:fig1} (d). During the gradient exposure process, some steps are produced on the arch bridge surface, as shown in Fig.~\ref{fig:fig1} (c). The deposited aluminum film should be thick enough to ensure that the aluminum film does not break during these steps. On the other hand, if the deposition layer is too thick, the aluminum film on top of airbridges will connect with that on the unexposed photoresist, and make it difficult for liftoff. Therefore, a moderate aluminum film thickness is critical for successfully fabricating airbridges using the gradient exposure technique.

Following aluminum film deposition, we immerse the wafer in an $80\ ^{\circ}\rm C$ N-Methylpyrrolidone (NMP) bath to simultaneously carry out the liftoff process and the release process. Compared to the re-flow fabrication technique, the liftoff process is much easier to implement due to the existence of high steps between the top of the airbridge and the unexposed photoresist. After the liftoff process, we obtain the arch-bridge-shaped airbridges in superconducting quantum circuits, as shown in Fig.~\ref{fig:fig1} (e). In Fig.~\ref{fig:fig1} (f), we present a false-colored scanning electron microscope (SEM) photograph of such an airbridge sample.

\begin{figure*}
\includegraphics[width=0.7\linewidth]{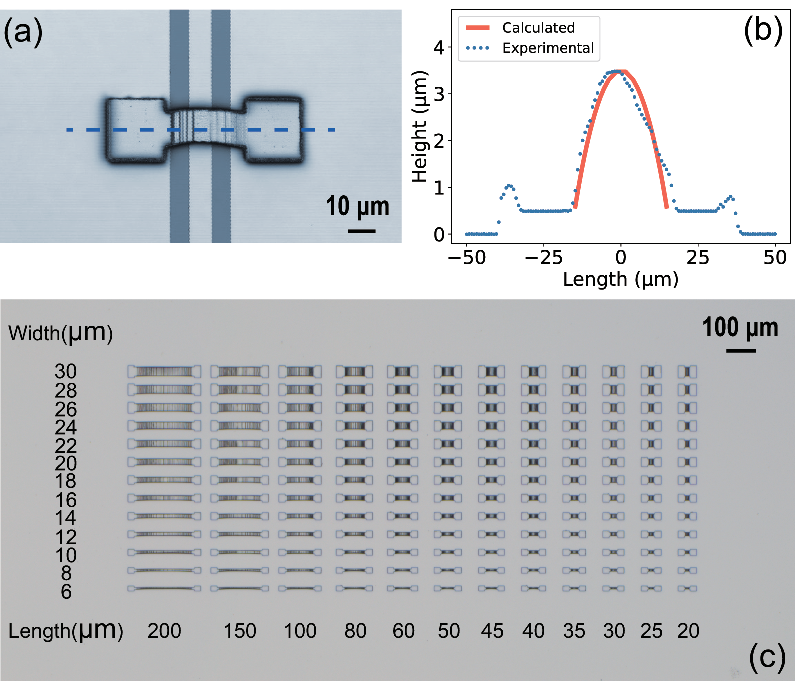}
\caption{\label{fig:fig2}(a) LSCM image of the airbridge fabricated using the gradient exposure technique. (b) The profile of the obtained airbridge. Blue dots represent experimentally measured data, and the orange line represents designed values. The height of bridge piers is about $\rm 500\ nm$, which is decided by the deposited Al thickness. (c) Optical micrograph of the bridges with different scales. The length of the airbridge is from $20\ \rm \mu m$ to $200\ \rm \mu m$, and the width is from $6\ \rm \mu m$ to $30\ \rm \mu m$.}
\end{figure*}

The profiles of the fabricated airbridges that are $30\ \rm \mu m$ long and $12\ \rm \mu m$ wide are examined with a laser scanning confocal microscope (LSCM),a Zeiss LSM 800, and the result is presented in Fig.~\ref{fig:fig2} (a). In this figure, steps in the bridge deck and a slight asymmetry of the airbridge relative to the CPW can be observed. The slight asymmetry is induced by the deviation in the alignment process and the resulted exposure imperfectness. To further and more precisely explore the obtained shape of the airbridge, we use a KLA-Tencor P-17 /P-7 Profiler to measure the profile of the airbridge. In Fig.~\ref{fig:fig2}(b), the profiles of the measured airbridge and the designed one are plotted. As can be seen, the profile of the airbridge deck is generally consistent with that of the designed profile, which is also consistent with that obtained with the LSCM. The small warping edges on the outer side of the piers are a result of the aluminum deposition on the photoresist side walls of the resist.
It is difficult to completely avoid sidewall deposition in the aluminum evaporation process. In our scheme, the thickness of the resist is up to $4\ \rm \mu m$.
Airbridges with flexible dimensions can be conveniently obtained using our technique. In Fig.~\ref{fig:fig2} (c), we present the fabricated airbridges with lengths from $20\ \rm \mu m$ to $200\ \rm \mu m$ and widths from $6\ \rm \mu m$ to $30\ \rm \mu m$. Based on the optical profile inspection, we can see that the structures of all airbridges are complete, and the arch bridge profiles with different dimensions are all preserved. The result indicates the high reliability of our technique.

With our technique, the airbridge profile is determined by the gradient exposure process. To obtain the proper arch bridge shape, we need to precisely control the applied exposure dosage in different areas. Therefore, it is necessary to first calibrate the relationship between exposure dosage and residual photoresist thickness with different substrate materials. In our process, the exposure dosage is determined by the applied laser power, and Beer's law~\cite{beer} is employed to estimate the thickness of residual photoresist (see the supplementary material for details)
\begin{equation}\label{eq1}
	z=z_0+\frac{1}{\alpha}\ln(\frac{P_{z_0-z}}{P-P_0}),
\end{equation}
\noindent where $z$ is the thickness of residual photoresist; $z_0$ is the initial thickness of the photoresist; $\alpha$ is the absorption coefficient of the photoresist, which depends on the property of the photoresist,~\cite{absorb} and is a constant for a certain laser wavelength; $P$ is the applied laser power; $P_0$ is the offset between the applied power and the power at the photoresist surface; and $P_{z_0-z}$ is the dose-to-clear, which specifies the required minimum exposure energy to remove the photoresist in the development process.~\cite{microchemicals_photoresist,photoresist_IEEEthesis}

\begin{figure*}[hptb]
\includegraphics[width=0.7\linewidth]{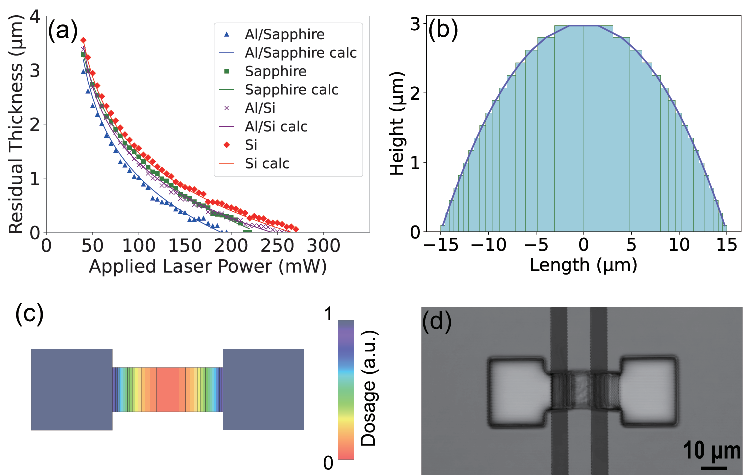}
\caption{\label{fig:fig3}
(a) Residual Thickness as a function of applied laser power with different substrates: the aluminum covered sapphire substrate (Al/sapphire substrate), the sapphire substrate, the aluminum covered silicon substrate (the Al/Si substrate) and the Si substrate.
The dots are experiment data, and the lines are calculated using Eq. \ref{eq1}, which show that residual thickness decreases exponentially with increasing laser power.
(b) The designed profile of the airbridge deck. An arc with certain height and length is used to depict the airbridge profile. Meanwhile, the required exposure laser power of each step is obtained with the fitted curve in Fig.~\ref{fig:fig2}(a).
(c) The designed exposure dosage of a airbridge. The colors indicate the different exposure dosages.
(d) LSCM image of the photoresist after gradient exposure and development.
}
\end{figure*}

Based on the Beer’s law, we can apply different laser power in the exposure process to realize different residual photoresist thickness. Taking into account that substrate materials can affect exposure results because of their different reflectivity, we perform the dose tests on four commonly used substrates: the aluminum-covered sapphire substrate (Al/sapphire substrate), the sapphire substrate, the aluminum-covered silicon substrate (the Al/Si substrate), and the Si substrate. After development, the residual photoresist thickness is measured by a KLA-Tencor P-17 /P-7 Profiler. The residual thickness of the photoresist with different exposure laser power is calculated using Eq.~\ref{eq1}. As shown in Fig.~\ref{fig:fig3} (a), the residual thickness exhibits an exponential decrease and asymptotically tends to zero with increasing laser power. For all substrate materials, the experiment results fit well with the calculated data obtained in Eq.~\ref{eq1}. Therefore, the exposure process can be properly described in Eq.~\ref{eq1}, which can be inversely used to calculate the required exposure laser power.

In order to obtain the proper airbridge profile, we need to first calibrate the exposure laser power. As shown in Fig.~\ref{fig:fig3}(b), the objective cross-section of the airbridge is depicted as an arc and illustrated in a Cartesian coordinate system, with the x-axis being the airbridge length and the y-axis being the airbridge height.
In this figure, the length and height of the bridge profile are set as $30\ \rm \mu m$ and $3\ \rm \mu m$, respectively.
Then, the height is divided into several steps by a fixed step height. It is worth noting that the step height needs to be smaller than the deposited metal thickness of the airbridge to ensure the integrity of the airbridge in the liftoff process. The more steps that are divided, the smoother the airbridge profile, with the trade-off being the longer exposure time. In our example, the airbridge height is divided into 18 steps, taking into consideration exposure accuracy and exposure time, as shown in Fig.~\ref{fig:fig3} (b-c).

To properly carry out the exposure procedure, we estimate the required laser power of each airbridge step in Eq.~\ref{eq1} and the parameters extracted from the fitting results in Fig.~\ref{fig:fig3} (a). The CPW transmission line is composed of two aluminum ground planes and an aluminum centerline, which are separated by two gaps. It is necessary to pay attention to the different exposure laser power required by different underlying materials. In Fig.~\ref{fig:fig3} (c), we present the designed exposure laser power pattern according to the results calculated in Eq.~\ref{eq1}. In this scheme, the airbridge is divided into 18 parts and patterned separately in 18 layers of an exposure file. During the exposure process, different exposure laser powers are assigned to their corresponding layers. In the bridge piers area, the applied laser power should be slightly larger than the calculated value to ensure that all resist can be removed in the development process. The layouts are generated in the GDSpy~\cite{gdspy} program with design parameters. After development, we obtain the arch bridge profile in the photoresist, as shown in Fig.~\ref{fig:fig3} (d) the LSCM image of the exposed photoresist.

\begin{figure}[hptb]
\includegraphics[width=\linewidth]{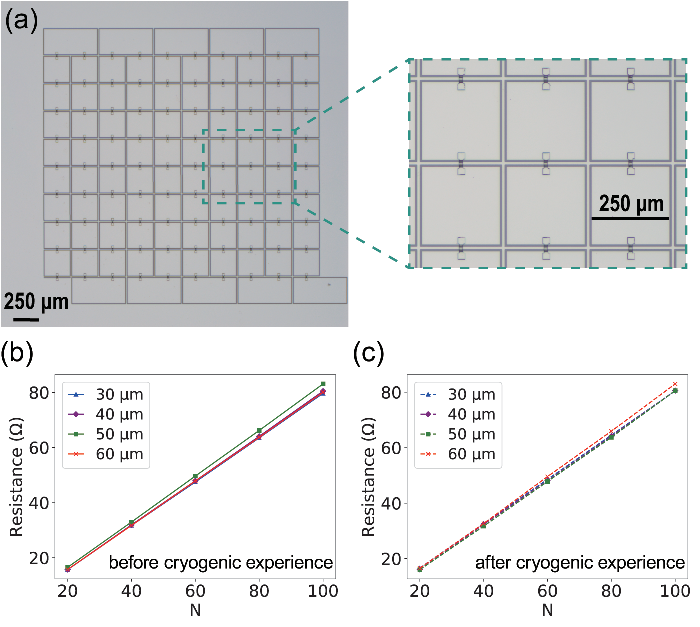}
\caption{\label{fig:fig4}
(a) Optical micrograph of 100 airbridges in a series. (b) Airbridge resistances before the cryogenic experience. (c) Airbridge resistances after the cryogenic experience.
}
\end{figure}

To estimate the reliability and the yield rate of our fabrication technique, we perform the room temperature resistance test on airbridges with a width of $12\ \rm \mu m$ and a length of $30\ \rm \mu m$, $40\ \rm \mu m$, $50\ \rm \mu m$ and $60\ \rm \mu m$. For each airbridge dimension, 100 airbridges above CPWs are connected in a series, as shown in Fig.~\ref{fig:fig4} (a). The room-temperature resistances of different numbers of bridges are measured with the four-probe method, and the measured resistance vs the number of airbridges in series $N$ are shown in Fig.~\ref{fig:fig4} (b). The measured resistance of each size of airbridges is linearly related to the airbridge number. The linear relationship indicates good uniformity of the airbridges produced by the gradient exposure technique.
Then, the measured samples are placed in a dilution refrigerator, and are cooled down to $10 \rm mK$. After the airbridges warm up to room temperature, the resistances of the series are measured again, and the results are shown in Fig.~\ref{fig:fig4}(d). A linear relationship as well as slight resistance variations are observed. These test results indicate that the fabricated airbridges have good robustness under the room temperature cryogenic temperature thermal circle. At the same time, the room temperature resistance of 36 such samples is tested (without cryogenic experience). Together with the optical profile inspection, the resistance tests show that all 3600 airbridges are well preserved, indicating a high yield rate of the gradient exposure airbridge fabrication technique.

\begin{figure}
\includegraphics[width=\linewidth]{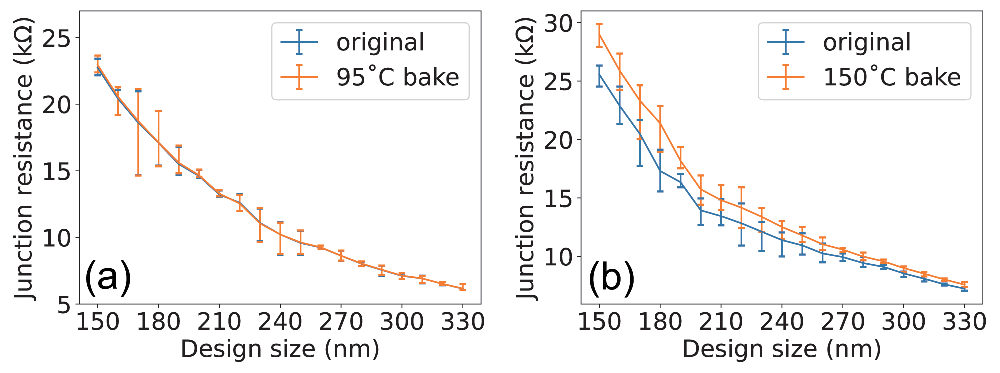}
\caption{\label{fig:fig5}
The measured Josephson junction resistances before and after the baking process. (a) Resistances of Josephson junctions with different junction areas before and after the $\rm 2mins\ 30s$ \ $95^{\ \circ} \rm C$ baking. (b) Resistances of Josephson junctions with different junction area before and after the $\rm 5mins$ $150^{\ \circ} \rm C$ baking.
}
\end{figure}

In most superconducting quantum circuits, the energy spectra of qubits depend on the critical current of Josephson junctions, which can be calculated based on their room temperature resistances.~\cite{JJ_res} Here, we estimate the effects of different airbridge fabrication processes on the critical current of Josephson junctions by measuring room temperature resistance variations. In our process, Megaposit SPR-220-3 photoresist coated samples need to be baked at $95^{\ \circ} \rm C$ for $2\rm mins \ 30\rm s$. In comparison, the re-flow followed lithography technique requires that the Megaposit SPR-220-3 photoresist coated samples be baked at around $150^{\ \circ} \rm C$ for $5\rm mins$ to form the arch bridge profile. To compare the effects of these two fabrication techniques, we fabricate Josephson junctions with junction areas from $150\times 150\ \rm nm^2 $ to $330\times 330 \ \rm nm^2$ on two single-side polished sapphire wafers using the bridge-free Manhattan style method~\cite{Manhattan}.
The room temperature resistances of Josephson junctions on each wafer are measured using the four-probe method. After the measurement, $4\ \rm\mu m$-thick Megaposit SPR-220-3 photoresist is coated on the two wafers. Then, one wafer is baked at $95^{\ \circ} \rm C$ for $2\rm mins \ 30\rm s$, while the other is baked at $150^{\ \circ} \rm C$ for $5\rm mins$. After clearing the photoresist, we measured the room temperature resistances of Josephson junctions on each wafer again.
The results are shown in Fig.~\ref{fig:fig5}. The $150^{\ \circ} \rm C$ baking process required by the re-flow followed lithography technique results in an approximately $\rm 10\%$ increase in junction resistances. In contrast, there is no significant change in junction resistances after the $95^{\ \circ} \rm C$ baking process. Therefore, our airbridge fabrication process will not significantly change the spectra of superconducting qubits.

\begin{figure}
\includegraphics[width=\linewidth]{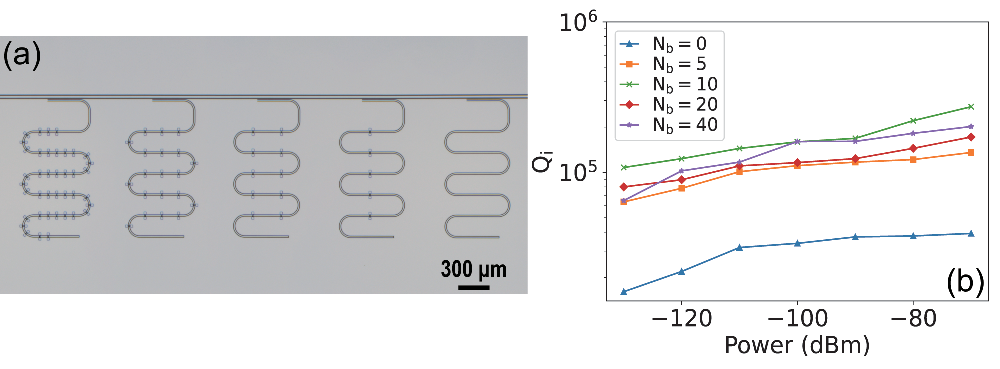}
\caption{\label{fig:fig6}
(a) Optical micrograph of five $\rm \lambda/4$ resonators with different numbers of airbridges coupled to the readout line. The number of airbridges $\rm N_b$ on the resonators is 0, 5, 10, 20, and 40 respectively. (b) The quality factor of resonators versus microwave power.
}
\end{figure}

Airbridges between discontinuous CPW grounds can eliminate parasitic modes in superconducting circuits; however, they may introduce dielectric loss into the circuits. Therefore, we characterize the effects of airbridges on the quality factors (Q factors) of CPW transmission line resonators.
Different numbers of airbridges are uniformly distributed on the resonators, as shown in Fig.~\ref{fig:fig6} (a). Then, the samples are cooled down to $10\ \rm mK$ in a dilution refrigerator, and the internal quality factor $\rm Q_i$ of the resonators with different input microwave powers is measured~\cite{curvefit}.  As shown in Fig.~\ref{fig:fig6} (b), the low power internal quality factor $\rm Q_i$ of the resonator without an airbridge is extremely low, which may result from the coupling between the resonator and parasitic modes. When the fabricated bridge quantity is less than 10, the low power $\rm Q_i$ increases with the increasing number of fabricated airbridges, indicating that superconducting airbridges are beneficial for eliminating parasitic modes in CPW circuits. However, airbridges may introduce additional two-level-system defects into the circuits, and thus reduce the low power quality factor $\rm Q_i$. With more airbridges above the resonators, the quality factor $\rm Q_i$ of the resonators decrease as the bridge number increases. Using these data, we can roughly estimate that each bridge introduces about $2\times 10^{-7}$ to the loss tangent of the resonator. This figure is about an order of magnitude larger than that in Ref [7]. The relatively large loss introduced by each bridge is likely caused by the poor quality of our aluminum film. The quality factors $\rm Q_i$ we measured is approximately one order of magnitude smaller than that in Ref [7].

In conclusion, we demonstrated superconducting airbridge fabrication technology based on the gradient exposure process. Using this technique, the arch bridge profile can be obtained by properly applying and gradually varying exposure dosage. Compared with the re-flow followed fabrication method,~\cite{chen2014fabrication} our technique reduced the process complexity and the negative impact on qubits.
Our fabrication technique does not require high temperature baking that can change the critical current of Josephson junctions and lead to variations in qubit spectra.~\cite{anneal}
The gradient exposure technique provides the ability to fabricate superconducting airbridges with other shapes and dimensions than the arch bridge profile presented above. The consistent electrical performance under thermal cycles shows the thermal robustness of the airbridges produced using our technique. A considerably high fabrication yield rate is also obtained. Our work provides a potential airbridge fabrication technique compatible with large-scale and high-coherence superconducting quantum circuits.

\section*{supplementary material}
See the supplementary material for the exposure kinetics of photoresist.

\begin{acknowledgments}
This work was partly supported by the Key R\&D Program of Guangdong Province (Grant No. 2018B030326001), NSFC (Grant Nos. 12074179, 11890704, and U21A20436), NSF of Jiangsu Province (Grant No. BE2021015-1), and the Innovation Program for Quantum Science and Technology (No. 2021ZD0301700).
\end{acknowledgments}

\section*{Data Availability Statement}
The data that support the findings of this study are available from the corresponding author upon reasonable request.

\section*{REFERENCES}
\bibliography{GEref}

\end{document}


\title{Supplementary Materials for \\
Fabrication of Airbridges with Gradient Exposure}

\author{Yuting Sun}
\altaffiliation{These authors contributed equally to this work}
\affiliation{National Laboratory of Solid State Microstructures, School of Physics, Nanjing University, Nanjing 210093, China}
\affiliation{Hefei National Laboratory, Hefei 230088, China}
\email[]{qu_syt@163.com}
\author{Jiayu Ding}
\altaffiliation{These authors contributed equally to this work}
\affiliation{National Laboratory of Solid State Microstructures, School of Physics, Nanjing University, Nanjing 210093, China}
\affiliation{Hefei National Laboratory, Hefei 230088, China}
\author{Xiaoyu Xia}
\affiliation{National Laboratory of Solid State Microstructures, School of Physics, Nanjing University, Nanjing 210093, China}
\affiliation{Hefei National Laboratory, Hefei 230088, China}
\author{Xiaohan Wang}%
\affiliation{National Laboratory of Solid State Microstructures, School of Physics, Nanjing University, Nanjing 210093, China}
\affiliation{Hefei National Laboratory, Hefei 230088, China}
\author{Jianwen Xu}
\affiliation{National Laboratory of Solid State Microstructures, School of Physics, Nanjing University, Nanjing 210093, China}
\affiliation{Hefei National Laboratory, Hefei 230088, China}
\author{Shuqing Song}
\affiliation{National Laboratory of Solid State Microstructures, School of Physics, Nanjing University, Nanjing 210093, China}
\affiliation{Hefei National Laboratory, Hefei 230088, China}
\author{Dong Lan}
\affiliation{National Laboratory of Solid State Microstructures, School of Physics, Nanjing University, Nanjing 210093, China}
\affiliation{Hefei National Laboratory, Hefei 230088, China}
\author{Jie Zhao}
\email[]{jiezhao@nju.edu.cn}
\affiliation{National Laboratory of Solid State Microstructures, School of Physics, Nanjing University, Nanjing 210093, China}
\affiliation{Hefei National Laboratory, Hefei 230088, China}
\affiliation{Shishan Laboratory, Suzhou Campus of Nanjing University, Suzhou 215000, China}
\author{Yang Yu}
\affiliation{National Laboratory of Solid State Microstructures, School of Physics, Nanjing University, Nanjing 210093, China}
\affiliation{Hefei National Laboratory, Hefei 230088, China}

\maketitle
\date{August 1 2022}

\begin{center}
	\textbf{EXPOSURE KINETICS}
\end{center}

In order to obtain the precise exposure dosage of a certain residual photoresist thickness, we need to understand the photoresist exposure process.
According to the Beer's Law, the light intensity in a positive photoresist can be described as:
\renewcommand{\theequation}{S\arabic{equation}}
\begin{equation}\label{beer}
	P(z^{\prime})=P(0)\exp(-\alpha z^{\prime}).
\end{equation}
where $P(z^{\prime})$ is the intensity at depth $z^{\prime}$; $P(0)$ is the intensity at the photoresist surface; and $\alpha$ is the absorption coefficient of the photoresist, which is a constant for a certain laser wavelength.

The dose at depth $z^{\prime}$ is defined by the product of the intensity $P(z^{\prime})$ and the exposure time $t$. We define the dose-to-clear $E_0=P_{z^{\prime}}\cdot t$, which represents the minimum energy required to remove the photoresist above $z^{\prime}$ during development. In this equation, $P_{z^{\prime}}$ is the corresponding critical intensity.

The photoresist thickness reduction $z^{\prime}$ can be expressed as,
\begin{equation}
    \begin{aligned}
        z^{\prime}=&-\frac{1}{\alpha}\ln(\frac{E_0}{P(0)\cdot t})\\
        =&-\frac{1}{\alpha}\ln(\frac{P_{z^{\prime}}}{P(0)})
	\end{aligned}
\end{equation}

Taking into account the offset between the power on the photoresist surface and the applied power,
\begin{equation}
	P(0)=P-P_0,
\end{equation}
where $P$ is the applied laser power, and $P_0$ is the power offset between $P$ and the power on the photoresist surface.
We obtain the equation:
\begin{equation}
    \begin{aligned}
	z^{\prime}&=-\frac{1}{\alpha}\ln(\frac{P_{z^{\prime}}}{P-P_0})
\\
	z&=z_0-z^{\prime}=-\frac{1}{\alpha}\ln(\frac{P_{z_0-z}}{P-P_0})
	\end{aligned}
\end{equation}
where $z_0$ is the initial thickness of the photoresist and $z$ is the thickness of residual photoresist.

\bibliography{main}